\documentclass[11pt]{article}
\usepackage[a4paper,margin=29mm]{geometry}
\usepackage{amsmath,amssymb,amsthm,booktabs,microtype}
\usepackage[numbers,square,comma]{natbib}
\usepackage[hidelinks]{hyperref}
\hypersetup{pdftitle={A deterministic O(n log n) algorithm for fixed-link shortest paths in concave Monge graphs},pdfauthor={Shucheng Chi and Lianyi Jia}}
\newtheorem{theorem}{Theorem}[section]
\newtheorem{lemma}[theorem]{Lemma}

\newtheorem{corollary}[theorem]{Corollary}
\newtheorem{definition}[theorem]{Definition}

\title{A deterministic \(O(n\log n)\) algorithm for fixed-link shortest paths in concave Monge graphs}
\author{Shucheng Chi\\[-1pt]
  {\small\href{mailto:chishuchengggg@gmail.com}{\texttt{chishuchengggg@gmail.com}}}
  \and
  Jiayi Lian\\[-1pt]
  {\small Zhejiang University}\\
  {\small\href{mailto:jiayilian@zju.edu.cn}{\texttt{jiayilian@zju.edu.cn}}}}
\date{}
\begin{document}
\maketitle
\begin{abstract}
\setcitestyle{authoryear,aysep={,}}
We give a deterministic \(O(n\log n)\)-time algorithm for finding a
minimum-weight source-to-sink path with exactly \(k\) edges in a concave Monge graph on \(n+1\) vertices. More precisely, the algorithm takes
\(O(n\log(h+2))\) time and \(O(n)\) space, where \(h=\min\{k-1,n-k\}\).
This resolves the 32-year-old open problem of finding an
\(n(\log n)^{O(1)}\)-time algorithm for every prescribed \(k\)~\citep{ast}.
The best previous worst-case bound was due to Schieber, whose algorithm
takes \(n\,2^{O(\sqrt{\log k\,\log\log n})}\) time
for \(k=\Omega(\log n)\)~\citep{schieber}. We give three applications.
\begin{itemize}
\setlength{\itemsep}{2pt}\setlength{\parsep}{0pt}
\item We give an \(O(n\log(h+2))\)-time algorithm for exact one-dimensional
\(k\)-means, which asks how to partition \(n\) sorted, positively weighted
points into \(k\) clusters to minimize the weighted sum of squared distances
to the cluster means. This answers the question of whether \(n\,\mathrm{polylog}(k)\)
time suffices~\citep{gronlund}. The classical dynamic program computes
the optima for all cluster counts through \(k\) in \(O(nk)\) time~\citep{wu}.
\item We give an \(O(n\log n)\)-time
algorithm for the inscribed \(k\)-gon problem, which asks us to choose
\(k\) of the cyclically ordered vertices of a convex \(n\)-gon to maximize
the perimeter of the polygon they form. The earlier matrix-search algorithm takes
\(O(nk+n\log n)\) time~\citep{smawk}. An \(O(n\log n)\) bound was known
for polynomially bounded integer coordinates with no three vertices
collinear~\citep{blp}.
\item For \(n\) sorted positive symbol weights and any feasible \(D\),
we construct an optimal tree in \(O(n\log(D+1))\) time for length-limited
binary prefix coding, which asks for a code that minimizes the weighted
sum of codeword lengths subject to a limit of \(D\) bits per codeword.
The package-merge algorithm takes \(O(nD)\) time~\citep{larmoreh}.
\end{itemize}
\end{abstract}

\section{Introduction}\label{sec:intro}

The \(k\)-link shortest-path problem asks for a shortest path among all paths from a given source to a given destination that
use exactly \(k\) edges. We study this problem on the ordered vertices
\(0,\ldots,n\), with a directed edge of weight \(w(i,j)\) from \(i\)
to \(j\) for every \(i<j\). The weights are finite real numbers that
can be queried in constant time. Specifically, we assume the graph satisfies the concave Monge condition.

\begin{definition}[Concave Monge graph]\label{def:monge}
A \emph{concave Monge graph} on \(0,\ldots,n\), where \(n\ge1\), has an edge of finite
real weight \(w(i,j)\) from \(i\) to \(j\) for every \(i<j\) and no
other edges. Its weights satisfy
\begin{equation}\label{eq:monge}
 w(a,c)+w(b,d)\le w(a,d)+w(b,c)
 \qquad(0\le a<b<c<d\le n).
\end{equation}
\end{definition}

We take \(0\) as the source and \(n\) as the destination. Given an integer
\(1\le k\le n\), the target is to find a path \(0=v_0<v_1<\cdots<v_k=n\) minimizing
\(\sum_{r=1}^k w(v_{r-1},v_r)\).

\subsection{Related work}\label{sec:related}

When the number of edges is unrestricted, a shortest path in a concave
Monge graph can be found in \(O(n)\) time~\cite{wilber,galil}. Fixing
the edge count introduces a difficulty. Two paths to the same vertex
may use different numbers of edges, so replacing a prefix by a cheaper
one can make the full path infeasible. The classical dynamic program
keeps a separate distance to each vertex for each edge count. Given the
distances for \(r-1\) edges, the SMAWK matrix-search algorithm finds a
best predecessor for every destination reachable in \(r\) edges in
\(O(n)\) total time~\cite{smawk,ast}. Thus the dynamic program takes
\(O(nk)\) time to answer a query for a single \(k\).

Bein, Larmore and Park~\cite{blp} approached the problem by adding the
same charge to every edge. A path with \(r\) edges then pays \(r\)
times the charge, and for some charge a shortest path uses exactly
\(k\) edges. For integer weights, they gave a search algorithm whose
running time depends on the magnitudes of the weights.
Aggarwal, Schieber and Tokuyama~\cite{ast} used parametric search to
obtain an \(O(n\sqrt{k\log n}+n\log n)\) bound for arbitrary real
weights. In 1994, they asked whether \(O(n\log^c n)\) time suffices for
every \(k\), for some constant \(c\).

Schieber~\cite{schieber} obtained a bound of
\(n\,2^{O(\sqrt{\log k\,\log\log n})}\) for \(k=\Omega(\log n)\)
by recursive parametric search, improving the bound when \(k=\Theta(n)\).
More recently, Wan~\cite{wan}
gave a linear-space algorithm whose running time decreases as \(k\)
approaches \(n\). Thus \(n(\log n)^{O(1)}\) time was known when
either \(k\) or \(n-k\) is at most polylogarithmic in \(n\). Whether
such a bound holds for every \(k\) remained open.

\subsection{Our contribution}\label{sec:contribution}

We give a deterministic \(O(n\log n)\)-time algorithm for every \(k\),
answering the question of Aggarwal, Schieber and Tokuyama with \(c=1\).
More precisely, the logarithmic factor depends only on the distance
from \(k\) to the nearer end of the range \(1,\ldots,n\).
Table~\ref{tab:bounds} compares the bounds for arbitrary real weights.

\begin{theorem}\label{thm:main}
Given a concave Monge graph on \(0,\ldots,n\) and an integer
\(1\le k\le n\), a deterministic algorithm finds a minimum-weight
path from \(0\) to \(n\) with exactly \(k\) edges and returns its weight
in \(O(n\log(h+2))\) time and \(O(n)\) space, including the output,
where \(h=\min\{k-1,n-k\}\). Each edge-weight query, exact arithmetic
operation and comparison is assumed to take constant time.
\end{theorem}

\begin{table}[!ht]
\centering\small
\caption{Deterministic time bounds for finding a shortest \(k\)-link path
on \(n+1\) vertices, where \(h=\min\{k-1,n-k\}\).}
\label{tab:bounds}
\begin{tabular}{@{}lll@{}}
\toprule
Method & Time bound & Worst case for \(k=\Theta(n)\)\\
\midrule
Layered dynamic program \cite{smawk,ast}
  & \(O(nk)\)
  & \(O(n^2)\)\\
Aggarwal et al.\ \cite{ast}
  & \(O(n\sqrt{k\log n}+n\log n)\)
  & \(O(n^{3/2}\sqrt{\log n})\)\\
Schieber \cite{schieber}
  & \(n\,2^{O(\sqrt{\log k\,\log\log n})}\)
  & \(n\,2^{O(\sqrt{\log n\,\log\log n})}\)\\
Wan \cite{wan}
  & \(O(\sqrt{nk(n-k+1)\log(n-k+1)})\)
  & \(O(n^{3/2}\sqrt{\log n})\)\\
Theorem~\ref{thm:main}
  & \(O(n\log(h+2))\)
  & \(O(n\log n)\)\\
\bottomrule
\end{tabular}
\end{table}

\subsection{Algorithm overview}\label{sec:overview}
Changing the edge count by one can move every interior vertex of an
optimal path. Our first observation is that, given any optimal path,
we can find an optimum with one more or one fewer edge in linear time
whenever that count is feasible. The two paths can be chosen to
\emph{interlace}. In vertex order, the \(j\)th interior vertex of the path
with more edges lies in the \(j\)th closed interval between consecutive
vertices of the other path. These intervals have linear total size, so
matrix search finds the new path in linear time (Lemma~\ref{lem:update}).
For inscribed polygons, Boyce et al.~\cite{boyce} proved a corresponding
interlacing property, and Aggarwal et al.~\cite[Section~6.1]{smawk}
used it to increase the number of sides by one.

Applying this update \(h\) times to the whole graph would take \(O(nh)\)
time. Instead, we divide the graph into \(h+1\) consecutive blocks of
nearly equal length and initially require the path to visit every block
boundary. We maintain a charge at which some shortest path through the
required boundaries has \(k\) edges. In each round we merge adjacent
blocks in pairs. Our second observation is that, at a fixed charge,
releasing one boundary changes the minimum and maximum edge counts of
the shortest paths by at most one each (Lemma~\ref{lem:release}).
If a round releases \(s\) boundaries, we can therefore restore the
prescribed count with at most \(s\) adjustments, each changing the
count in one block by one. With \(b\) blocks remaining, \(s\le b\)
and each block has length \(O(n/b)\). The path computations thus take
\(O(n)\) time per round. After \(O(\log(h+2))\) rounds only one block
remains, and its optimal path is the desired path.

\section{Preliminaries}\label{sec:prelim}

\paragraph{Notation.}
For vertices \(u<t\), let \(I=[u,t]\) denote the interval of vertices
\(u,u+1,\ldots,t\), and let \(m=t-u\) be its length. The graph induced
by \(I\) is again a concave Monge graph. We write \(w(P)\) for the
sum of the edge weights of a path \(P\).

\paragraph{Monge matrices and SMAWK.}
A matrix \(A\) is \emph{Monge} if for every pair of rows \(i<i'\) and
columns \(j<j'\),
\[
   A[i,j]+A[i',j']\le A[i,j']+A[i',j].
\]
In a Monge matrix, the column index of the leftmost minimum can only
increase or stay the same as we move down the rows. The same holds in
every submatrix. A matrix with this latter property is called
\emph{totally monotone} for row minima. Given constant-time access to
the entries of a totally monotone \(p\times q\) matrix, the SMAWK
algorithm finds the leftmost minimum in every row in \(O(p+q)\)
time~\cite{smawk}. It reduces the set of candidate columns, finds
the minima in every other row recursively, and uses these positions
to bound the searches in the remaining rows.

\paragraph{Shortest paths in concave Monge graphs.}
Fix a source \(u\) in a concave Monge graph, and let \(d(y)\) be the
minimum weight of a path from \(u\) to \(y\) with any number of edges.
A path to \(y>u\) ends with an edge \(x\to y\), so
\[
   d(u)=0,\qquad
   d(y)=\min_{u\le x<y}\bigl(d(x)+w(x,y)\bigr).
\]
Evaluating the minimum separately for each \(y\in[u,t]\) takes
\(O(m^2)\) time, where \(m=t-u\). For concave Monge weights, the
algorithms of Wilber and of Galil and Park compute all these distances
in \(O(m)\) time~\cite{wilber,galil}. The same bound holds after adding
a common constant to every edge weight, since both sides of
\eqref{eq:monge} change by the same amount.

\section{The pairwise merging algorithm}\label{sec:alg}

In this section we present the algorithm for Theorem \ref{thm:main}. Our goal is to find a charge \(c\) such that, after adding \(c\) to
every edge weight, some shortest path from \(0\) to \(n\) uses
exactly \(k\) edges.

\begin{definition}[Edge charge]\label{def:edge-charge}
Let \(G=(V,E,w)\) be the input graph. For \(c\in\mathbb R\), let
\(G_c=(V,E,w_c)\), where \(w_c(i,j)=w(i,j)+c\) for every edge
\(i\to j\). Thus a path \(P\) with \(r\) edges has weight
\(w_c(P)=w(P)+rc\) in \(G_c\).
\end{definition}

Since every \(k\)-edge path gains the same weight \(kc\), a shortest
path in \(G_c\) that uses \(k\) edges is a minimum-weight \(k\)-edge
path in \(G\). We can query an edge of \(G_c\) in constant time by
adding \(c\) to its weight in \(G\).

To find such a charge, we initially require the path to visit a set
of vertices that includes \(0\) and \(n\). For each required vertex
\(u<n\), let \(t\) be the next required vertex. We call the interval
\(I=[u,t]\) a \emph{block}. A path through all the required vertices
is shortest among such paths in \(G_c\) precisely when its portion in
each block is shortest between the endpoints of that block.

\begin{definition}[Cutoff]\label{def:cutoff}
Fix the required vertices. Minimize \(w_c(P)\) over all paths \(P\)
from \(0\) to \(n\) that visit every required vertex. We call \(c\)
a \emph{cutoff} for \(k\) if at least one minimizing path has exactly
\(k\) edges.

We use \(c=+\infty\) as shorthand for taking the direct edge in each
block, as one would for a sufficiently large finite charge.
\end{definition}

We start with a cutoff for the initial blocks. In each round we remove
boundaries and find a cutoff for the remaining blocks. When only
\(0\) and \(n\) are required, the cutoff applies to the original problem.

\subsection{Testing a cutoff}

Fix a partition and a finite charge \(c\). In each block \(I=[u,t]\),
consider all shortest paths from \(u\) to \(t\) in \(G_c\). Let
\(r^-_I\) be the fewest edges on any such path, and \(r^+_I\) the
most. Call paths attaining these counts the \emph{extreme paths}.
Adding a charge preserves the Monge inequality, so the following
results apply in \(G_c\).

\begin{lemma}[Extreme edge counts~{\cite[Theorem~3.1 and Lemma~3.4]{wan}}]\label{lem:extreme-paths}
The two extreme paths in \(I\) can be found in \(O(t-u)\) time and
space. Choosing the leftmost minimizing predecessor in the shortest-path
recurrence gives a path with the fewest edges, and choosing the
rightmost gives one with the most. For a fixed source both counts are
nondecreasing as the destination moves right. For a fixed destination
both are nondecreasing as the source moves left.
\end{lemma}

The fixed-destination statement follows by reversing the vertex order
and the edges.

\begin{lemma}[Path exchange~{\cite[Lemma~3.2]{wan}}]\label{lem:exchange}
Given the two extreme paths in \(I\), one can construct a shortest
\(u\)-to-\(t\) path in \(G_c\) with any prescribed count
\(r^-_I\le r\le r^+_I\) in \(O(r^+_I)\) time and space.
\end{lemma}

Let \(L=\sum_I r^-_I\) and \(U=\sum_I r^+_I\). The smallest possible
total for a shortest path through all required vertices is \(L\).
By Lemma~\ref{lem:exchange}, each block can use every count from
\(r^-_I\) to \(r^+_I\), so we can attain every total up to \(U\).
Thus \(c\) is a cutoff exactly when \(L\le k\le U\). Under the
direct-edge convention \(c=+\infty\), each block contributes one edge
to both totals.

\subsection{Initialization, merge rounds, and output}\label{sec:merge-round}

\paragraph{Initialization.}
If \(k=1\) or \(k=n\), return the unique feasible path and its weight.
Otherwise divide the \(n\) unit edges \(i\to i+1\) into
\(B=\min\{k,n-k+1\}=h+1\) consecutive blocks whose lengths differ
by at most one. If \(B=k\), the direct edges of the blocks already
give \(k\) edges, so \(+\infty\) is a cutoff. Otherwise
\(B=n-k+1\). Start with the path of all \(n\) unit edges. Apply the
removal rule in Section~\ref{sec:block-adjustment} \(n-k=B-1\)
times. The last removal gives the initial cutoff.

\paragraph{A merge round.}
Let \(c\) be the current cutoff. Pair consecutive blocks from left to
right and erase the boundary inside each pair. Carry an unpaired last
block forward. Removing a boundary allows paths to skip a formerly
required vertex. Keep \(c\) fixed and compute the extreme paths and
the totals \(L,U\) for the new blocks. There are three cases.

\begin{itemize}
\item If \(L\le k\le U\), retain \(c\). The test already guarantees
a shortest path through the remaining boundaries with \(k\) edges.
\item If \(U<k\), the old charge gives too few edges, even when we
choose the maximum count in every block. Start with these paths and
make \(k-U\) replacements, each increasing one block's count by one.
The new cutoff will be lower.
\item If \(k<L\), the old charge gives too many edges, even when we
choose the minimum count in every block. Start with these paths and
make \(L-k\) replacements, each decreasing one block's count by one.
The new cutoff will be higher.
\end{itemize}

For either direction, Section~\ref{sec:block-adjustment} specifies
which block to change and how the last replacement gives a new cutoff
\(c'\). Carry only the partition and its cutoff into the next round.
If the round erased \(s\) boundaries, each endpoint of the interval
of attainable counts moved by at most \(s\) at the old charge
(Lemma~\ref{lem:release}). The interval contained \(k\) before the
merge, so at most \(s\) replacements are needed.

\paragraph{Output.}
Repeat the merge round until only \([0,n]\) remains. Its cutoff \(c\)
is finite because \(k>1\) and the direct edge uses only one edge.
Find the extreme paths from \(0\) to \(n\) in \(G_c\). If either has
\(k\) edges, take it. Otherwise use Lemma~\ref{lem:exchange} to
obtain a shortest path \(P\) in \(G_c\) with exactly \(k\) edges.
Return the chosen path and its weight in \(G\).

\subsection{Choosing a replacement and the new cutoff}\label{sec:block-adjustment}

Suppose we need to increase or decrease the total edge count. Let \(P\)
be the current path in a block, and let \(r\) be its edge count.
At a finite inherited charge, the starting extreme path is optimal in
\(G\) among paths with \(r\) edges, since all such paths pay the same
charge. The direct edge and the path of all unit edges are each the
only path in the block with its edge count.
For each block that can change in the required direction, compute a
replacement \(Q\) of minimum weight in \(G\) with \(r+1\) or \(r-1\)
edges, respectively. Lemma~\ref{lem:update} finds \(Q\) in time
linear in the block length by searching for a path that interlaces \(P\).

\paragraph{Which block changes?}
Replacing \(P\) by \(Q\) changes the total weight in \(G\) by
\(w(Q)-w(P)\). Choose the candidate with the smallest such change.
Keep one candidate per eligible block in a min-heap keyed by this
difference, breaking ties by block order. For finite \(c\), the
ranking is the same in \(G_c\). All candidates add one edge or all
remove one edge, so their differences all acquire the same term
\(+c\) or \(-c\). Each heap entry stores a pointer to its candidate
path. Extract the minimum and adopt its path. If another replacement
is needed, only the block just chosen needs a new candidate, provided
it can change again in the same direction.

\paragraph{Which charge do we keep?}
Let \(P_r\) and \(P_{r+1}\) be the two paths in the last replacement,
indexed by their edge counts. In \(G_{c'}\) they tie exactly when
\[
   w(P_r)+r c'=w(P_{r+1})+(r+1)c',
   \qquad\text{that is,}\qquad
   c'=w(P_r)-w(P_{r+1}).
\]
Use this value as the new cutoff, whether the replacement added an
edge or removed one.

Why does a tie in one block give a cutoff for all the blocks? Let
\(F_I(r)\) be the minimum weight in \(G\) of an \(r\)-edge path across
\(I\). The charge at which counts \(r\) and \(r+1\) tie is precisely
the saving \(\Delta_I(r)=F_I(r)-F_I(r+1)\) from adding the edge.

\begin{lemma}[Discrete convexity~{\cite[Lemmas~2.2 and~2.3]{blp}}]\label{lem:block-counts}
In a block of length \(m\), the savings
\(\Delta_I(1),\ldots,\Delta_I(m-1)\) are nonincreasing. Consequently,
for finite \(c\), a shortest path across the block in \(G_c\) can
use exactly \(r\) edges if and only if \(c\) satisfies the applicable
bounds
\begin{equation}\label{eq:cutoff}
   \Delta_I(r-1)\ge c \quad(r>1),
   \qquad
   \Delta_I(r)\le c \quad(r<m).
\end{equation}
\end{lemma}

Using \(r\) edges selects the first \(r-1\) savings. On an addition,
the heap takes the largest available saving. On a removal, it gives
up the smallest selected saving. After the last replacement, every
selected saving is at least \(c'\) and every unselected saving is at
most \(c'\) (Section~\ref{sec:release-rounds}). By~\eqref{eq:cutoff}
the chosen paths are therefore shortest in their blocks in \(G_{c'}\).
Together they use \(k\) edges.

\paragraph{Example.}
Take \(k=4\) and \(w(i,j)=(x_j-x_i)^2\) for
\(x=(0,1,2,9,10,11,12,13,14)\). The four initial blocks have length
two. Their direct edges give the path \((0,2,4,6,8)\), written as
indices into \(x\). Table~\ref{tab:merge-example} shows what happens
when the three interior boundaries are removed over two rounds.

\begin{table}[!ht]
\centering\small
\caption{Each round tests the old cutoff \(c\) on the new blocks.
The last replacement determines \(c'\).}\label{tab:merge-example}
\begin{tabular}{@{}lcccc@{}}
\toprule
Blocks after merging & Old \(c\) & \([L,U]\) at old \(c\) & Edges to add & New \(c'\)\\
\midrule
\([0,4], [4,8]\) & \(+\infty\) & \([2,2]\) & \(2\) & \(14\)\\
\([0,8]\)        & \(14\)      & \([3,3]\) & \(1\) & \(12\)\\
\bottomrule
\end{tabular}
\end{table}

In the first round, two additions in \([0,4]\) save \(32\) and \(14\)
in \(G\), while the first available addition in \([4,8]\) would
save \(8\). Thus both additions go to the first block, and the new
cutoff is \(14\). In the second round the shortest paths in \(G_{14}\)
have three edges and weight \(78\) in \(G\). A best four-edge
replacement is \((0,2,3,5,8)\), of weight \(66\) in \(G\). The new
cutoff is \(78-66=12\). In \(G_{12}\) both paths have weight \(114\),
so the four-edge path attains the shortest-path distance.

\section{Correctness}\label{sec:correct}\label{sec:release-rounds}

A merge round starts with a charge \(c\) that works for the old
partition. We will show that erasing a boundary has a limited effect
at this same charge, and that each replacement made by the heap gives
a cutoff for the new total edge count. Together these facts prove the
following theorem. We then apply it inductively, starting with the
initial partition and ending with a single block.

\begin{theorem}[Restoring a cutoff]\label{thm:round}
Suppose a round starts with a cutoff for \(k\) and erases \(s\)
boundaries. It ends with a cutoff for \(k\) on the new partition and
makes at most \(s\) one-edge replacements.
\end{theorem}

We first justify the operation used to produce a heap candidate.
Given one optimal path, we need an optimal path with the adjacent
edge count. The two paths may share vertices.

\begin{definition}[Interlacing]\label{def:interlacing}
Paths \(S=(s_0,\ldots,s_q)\) and \(T=(t_0,\ldots,t_{q+1})\) with
the same endpoints \emph{interlace} if \(s_{j-1}\le t_j\le s_j\)
for \(1\le j\le q\).
\end{definition}

\begin{lemma}[Uncrossing two paths]\label{lem:rearrange}
Let \(A=(a_0,\ldots,a_p)\) and \(B=(b_0,\ldots,b_p)\) be paths with
the same number of edges. Form two vertex lists by taking, at each
position \(j\), the smaller and the larger of \(a_j,b_j\),
respectively. Both lists are paths, and their total weight is at most
\(w(A)+w(B)\).
\end{lemma}
\begin{proof}
Both lists are strictly increasing because \(A\) and \(B\) are.
Consider the two edges between positions \(j\) and \(j+1\). If their
tails and heads occur in the same order, taking minima and maxima
preserves the edges. If the order reverses, the original edges have
the form \(a\to d\) and \(b\to c\) with \(a\le b<c\le d\). The new
edges are \(a\to c\) and \(b\to d\). By~\eqref{eq:monge} their total
weight is at most the original total, with equality if the tails or
the heads coincide. Summing over the positions proves the claim.
\end{proof}

\begin{lemma}[An optimal neighbor of a given path]\label{lem:neighbor}
For two consecutive feasible counts \(q,q+1\) in a block, every path
optimal in \(G\) for either count interlaces a path optimal for the
other count.
\end{lemma}
\begin{proof}
Let \(S\) and \(T\) be optimal paths with \(q\) and \(q+1\) edges,
respectively. One is the given path. We apply Lemma~\ref{lem:rearrange}
twice, keeping the given path after each application.

First align the paths at their common source. Compare \(S\) with
\((t_0,\ldots,t_q)\), assigning the larger vertex at each position to
\(S\) and the smaller to \(T\). Since \(t_q<s_q=t_{q+1}\), the last
edge \(t_q\to t_{q+1}\) of \(T\) stays in place. The resulting two
paths have the original endpoints and edge counts, and the sum of
their weights has not increased. Each must therefore be optimal for
its count. Retain the new version of the other path and restore the
given one. If \(S\) was given, the new \(t_j\) is the smaller of
the two original vertices. If \(T\) was given, the new \(s_j\) is
the larger. Either choice gives \(t_j\le s_j\) for \(0\le j\le q\).

Next align the paths at their common destination. Compare \(S\) with
\((t_1,\ldots,t_{q+1})\), assigning the smaller vertex to \(S\) and
the larger to \(T\). The first edge of \(T\) stays in place since
\(s_0=t_0<t_1\). The same weight argument lets us keep the resulting
version of the other path while restoring the given one. Now
\(s_j\le t_{j+1}\) for \(0\le j\le q\). The first set of inequalities
also survives. If \(S\) changes, its new \(j\)th vertex is
\(\min(s_j,t_{j+1})\ge t_j\). If \(T\) changes, its new \(j\)th
interior vertex is \(\max(t_j,s_{j-1})\le s_j\). Together the two
sets of inequalities give interlacing.
\end{proof}

\begin{lemma}[One-edge update]\label{lem:update}
Given a path optimal in \(G\) for its edge count in a block of length
\(m\), we can find an optimal path and its weight for either feasible
adjacent count in \(O(m)\) time and space.
\end{lemma}
\begin{proof}
Write the given path as \(a_0,\ldots,a_r\). For an increase, use the
\(r\) intervals \([a_{j-1},a_j]\), \(1\le j\le r\), as the sets of
possible interior vertices of the new path, in order. For a decrease,
use \([a_j,a_{j+1}]\), \(1\le j\le r-2\). These intervals are the
\emph{layers} of the dynamic program. Remove the two block endpoints
from all interior layers and give each endpoint its own singleton
layer. By Lemma~\ref{lem:neighbor}, choosing one vertex from each
layer in strictly increasing order can produce an optimum for the
requested count.

Let \(D_i(y)\) be the minimum original weight of a path from \(a_0\)
to \(y\) that takes one vertex from each of layers \(0,\ldots,i\).
Start with \(D_0(a_0)=0\). Let \(X\) be the reachable vertices in
layer \(i-1\) and \(Y\) the vertices in layer \(i\), both in increasing
order. Taking the last edge from \(x\) to \(y\) gives
\begin{equation}\label{eq:layer-transition}
   D_i(y)=\min_{x\in X,\,x<y}\bigl(D_{i-1}(x)+w(x,y)\bigr)
   \qquad(y\in Y).
\end{equation}
An empty minimum means that \(y\) is unreachable. An interlacing
optimum exists, so every layer contains a reachable vertex.

The intervals are ordered, with \(\max X\le\min Y\). If the inequality
is strict, the matrix in~\eqref{eq:layer-transition}, with rows \(y\)
and columns \(x\), is a full Monge matrix. Adding \(D_{i-1}(x)\) to
column \(x\) preserves~\eqref{eq:monge}. If \(\max X=\min Y=z\),
scan row \(z\) over \(x<z\), leaving a full matrix on the remaining
rows. SMAWK on the full matrix and the possible extra scan find all
distances and their minimizing predecessors in \(O(|X|+|Y|)\) time.
Each vertex belongs to at most two layers, so their total size is
\(O(m)\). Following the stored predecessors from \(a_r\) recovers
the path within the same time and space bounds.
\end{proof}

We next bound how far the target can be from the new interval of
attainable counts. Here the charge stays fixed while a boundary is
erased.

\begin{lemma}[Releasing one boundary]\label{lem:release}
Erasing one boundary changes each endpoint of \([L,U]\) by at most
one at any fixed charge, including the direct-edge convention
\(c=+\infty\).
\end{lemma}
\begin{proof}
Suppose \([u,v]\) and \([v,t]\) merge. At a finite charge \(c\), let
\(e(x,y)\) denote the minimum number of edges on a shortest path from
\(x\) to \(y\) in \(G_c\), with \(e(x,x)=0\). The same argument applies
throughout with the maximum in place of the minimum. We compare the
old contribution \(e(u,v)+e(v,t)\) with the new one \(e(u,t)\).

Choose a shortest \(u\)-to-\(t\) path with \(e(u,t)\) edges. Each
subpath is shortest and attains the same choice of extreme count for
its endpoints. Otherwise replacing the subpath would improve either
the weight of the whole path or its edge count among shortest paths.
If the path visits \(v\), the two contributions are equal.

Otherwise the path has an edge \(x\to y\) with \(x<v<y\). Suppose it
has \(i\) edges before \(x\) and \(j\) edges after \(y\). Its subpaths
give \(e(u,x)=i\), \(e(u,y)=i+1\), \(e(x,t)=j+1\), and \(e(y,t)=j\).
By Lemma~\ref{lem:extreme-paths}, first fixing \(u\) and then fixing
\(t\), we have
\[
 i\le e(u,v)\le i+1,\qquad j\le e(v,t)\le j+1 .
\]
The old contribution thus differs from \(e(u,t)=i+1+j\) by at most
one. All other blocks keep their contributions. For \(c=+\infty\),
two direct edges become one, so both endpoints decrease by one.
\end{proof}

It remains to justify the new charge. A block path with \(r\) edges
selects the first \(r-1\) entries of the list
\(\Delta_I(1),\ldots,\Delta_I(m-1)\). These entries are
nonincreasing by Lemma~\ref{lem:block-counts}. We distinguish equal
entries by their positions in their respective lists.

\begin{lemma}[The charge after a replacement]\label{lem:heap-cutoff}
Suppose each block path is optimal in \(G\) for its count, and every
selected saving is at least every unselected saving. After one feasible
heap replacement, let \(v\) be the saving selected or discarded.
At charge \(v\), each resulting block path is shortest in \(G_v\).
The ordering of selected and unselected savings is preserved.
\end{lemma}
\begin{proof}
For an addition in a block with \(r\) edges, the candidate from
Lemma~\ref{lem:update} has heap key
\(F_I(r+1)-F_I(r)=-\Delta_I(r)\). The first unselected saving in
each block is its largest unselected saving. The heap therefore takes
a largest unselected saving overall. For a removal, the key is
\(F_I(r-1)-F_I(r)=\Delta_I(r-1)\), and the heap discards a smallest
selected saving overall. In both cases all entries selected after
the operation are at least \(v\), and all entries left unselected
are at most \(v\). This includes ties at \(v\).
For each block these are precisely the applicable inequalities
in~\eqref{eq:cutoff}. The chosen path is optimal for its count in
\(G\), so it is a shortest path in \(G_v\).
\end{proof}

\begin{proof}[Proof of Theorem~\ref{thm:round}]
At the inherited charge \(c\), the old interval contained \(k\).
Apply Lemma~\ref{lem:release} to the \(s\) boundaries in succession,
always at this charge. Each endpoint moves by at most \(s\) in total.
Write \([L,U]\) for the resulting interval. If \(L\le k\le U\), the
old charge remains a cutoff. Otherwise the algorithm makes exactly
\(k-U\) additions or \(L-k\) removals. In either case the gap is
at most \(s\). Each requested step is feasible because
the target lies between the number of remaining blocks and \(n\).

For finite \(c\), an increase from \(r\) to \(r+1\) changes the
best weight in \(G_c\) by \(c-\Delta_I(r)\). The maximum optimal count
therefore selects all savings at least \(c\), and the minimum optimal
count selects exactly those strictly greater than \(c\). Thus in
either adjustment case the starting allocation satisfies the ordering
in Lemma~\ref{lem:heap-cutoff}. At \(c=+\infty\) the algorithm starts
with the direct edges and the selected set is empty.

Apply Lemma~\ref{lem:heap-cutoff} at each replacement. On the last
one its saving \(v\) equals the difference in \(G\) between the
weights of the paths with fewer and more edges, precisely the new
charge \(c'\) chosen in Section~\ref{sec:block-adjustment}. The block
paths are shortest in \(G_{c'}\) and their counts sum to \(k\).
Concatenating them witnesses the new cutoff. For finite old \(c\),
all added savings were strictly below \(c\), and all discarded savings
in a removal round were strictly above \(c\). This also gives the
direction of the charge change asserted in the algorithm.
\end{proof}

\begin{theorem}[Correctness of the algorithm]\label{thm:correct-output}
For every \(1\le k\le n\), the pairwise merging algorithm returns a
minimum-weight \(k\)-edge path from \(0\) to \(n\) in \(G\) and its
weight.
\end{theorem}
\begin{proof}
The cases \(k=1\) and \(k=n\) have unique feasible paths. For
\(1<k<n\), if \(B=k\), the \(B\) direct edges witness a cutoff under
the \(+\infty\) convention. Otherwise \(B=n-k+1\). The path of all
unit edges is optimal for its count and selects every saving. After
\(n-k=B-1\) removals, Lemma~\ref{lem:heap-cutoff} makes the last
discarded saving a cutoff for \(k\).

By Theorem~\ref{thm:round}, every round preserves the existence of a
cutoff for \(k\). The final block \([0,n]\) has a finite cutoff \(c\),
since its direct edge has count one. The extreme paths in \(G_c\)
bracket \(k\), and Lemma~\ref{lem:exchange} gives a shortest path
\(P\) with exactly \(k\) edges, including the case that an extreme
path already has this count. Every \(k\)-edge path pays the same
charge \(kc\), so \(P\) has minimum weight in \(G\). The algorithm
returns \(w(P)\).
\end{proof}

For interior \(k\) the final cutoff also certifies the result. Given
\(P\) and \(c\), check its endpoints, its \(k\) forward edges and
their original weight. Then compare \(w(P)+kc\) with the unrestricted
shortest-path distance from \(0\) to \(n\) in \(G_c\). Equality
certifies optimality, and both checks take \(O(n)\) time.

\section{Time and space analysis}\label{sec:time}

There are two costs to account for in a round. Computing a new
candidate path takes time proportional to the length of its block.
Updating the heap takes time logarithmic in the number of blocks.
We bound the path computations within each round and count the heap
updates over all rounds.

Assume \(1<k<n\), and recall that \(B=\min\{k,n-k+1\}=h+1\).
Number the initial partition as level zero. At level \(\ell\) there
are \(b=\lceil B/2^\ell\rceil\) blocks. Each contains at most
\(2^\ell\) initial blocks of length at most \(\lceil n/B\rceil\).
Writing \(M\) for the longest block at this level, we obtain
\[
 bM \le (B+2^\ell)\lceil n/B\rceil \le 6n .
\]
Indeed, \(\ell\le\lceil\log_2 B\rceil\) gives \(2^\ell<2B\),
and \(B\lceil n/B\rceil\le n+B\le 2n\).

The block lengths at a level sum to \(n\). Finding both extreme
paths and the first candidate in each eligible block therefore takes
\(O(n)\) time in total by Lemmas~\ref{lem:extreme-paths}
and~\ref{lem:update}. Building the heap from these candidates takes
\(O(b)\) time. A round that ends with \(b\) blocks erases at most
\(b\) boundaries, one for each merged pair.
Theorem~\ref{thm:round} bounds the number of replacements by the
number of erased boundaries. After a replacement only the chosen
block needs a new candidate, which takes \(O(M)\) time. Thus even
if every replacement occurs in the longest block, all path
computations at this level take \(O(n+bM)=O(n)\) time. The same
bound applies at level zero, where initialization makes at most
\(B-1\) replacements.

Over all merge rounds, exactly \(B-1\) boundaries are erased. Thus
there are at most \(B-1\) heap extractions in these rounds and at
most \(B-1\) more in initialization. Each extraction and possible
reinsertion takes \(O(\log(B+1))\) time in the heap. The
\(1+\lceil\log_2 B\rceil\) levels therefore take
\[
 O\bigl(n(1+\lceil\log_2 B\rceil)+B\log(B+1)\bigr)
     = O(n\log(h+2))
\]
time. At the final cutoff, computing the extreme paths, recovering
the requested path by Lemma~\ref{lem:exchange}, and summing its
original edge weights take a further \(O(n)\) time.

In a block of length \(m\), the extreme paths, the current path, a
candidate path and the predecessor records occupy \(O(m)\) words.
Heap entries point to paths, so moving an entry costs constant time.
The edge weights are queried on demand, and the algorithm can
release a level's paths when it passes its cutoff to the next level.
Since the block lengths sum to \(n\), the algorithm uses \(O(n)\)
space, including the output. For \(k=1\) or \(k=n\), returning
the unique path and its weight takes at most \(O(n)\) time and
space. Together with Theorem~\ref{thm:correct-output}, this proves
Theorem~\ref{thm:main}.

\section{Applications}\label{sec:applications}

In each reduction a prescribed number of clusters, polygon sides, or
tree levels becomes a prescribed number of edges. We describe how to
evaluate the edge weights and how to recover a solution from the
resulting path.

\subsection{Exact one-dimensional clustering and scalar quantization}

In weighted one-dimensional \(k\)-means, the input consists of \(n\)
observations \(x_i\in\mathbb R\) with positive weights \(p_i\).
We seek a partition into \(k\) nonempty clusters and a centre for
each cluster that minimize the sum of weighted squared distances to
the assigned centres. In scalar quantization the centres are the
\(k\) reconstruction values, and the objective is the squared
reconstruction error. In unweighted \(k\)-medians the objective is
the sum of absolute distances to the assigned centres.

On sorted input, Wu's algorithm for scalar quantization computes the
optimum for every cluster count up to \(k\) in \(O(nk)\)
time~\cite{wu}. For distinct integer coordinates in \([1,X]\) and
integer weights in \([1,W]\), Bein, Larmore, and Park obtained an
\(O(n(1+\log(nWX)))\) bound~\cite[Section~3.3]{blp}.
For a single prescribed \(k=\Omega(\log n)\), Schieber's algorithm
gives \(n\,2^{O(\sqrt{\log k\,\log\log n})}\)
time~\cite{schieber,gronlund}. Gr{\o}nlund et al.\ asked whether a
single prescribed \(k\) could be handled in \(n(\log n)^{O(1)}\) time,
or even \(n(\log k)^{O(1)}\) time on sorted
input~\cite[Final remarks]{gronlund}.

Theorem~\ref{thm:main} answers the latter question with a single
logarithmic factor, also for positive real weights. The factor
decreases again as \(k\) approaches \(n\). For unsorted input,
sorting attains the worst-case bound in the following corollary.

\begin{corollary}\label{cor:means}
Let \(1\le k\le n\) and \(h=\min\{k-1,n-k\}\). An optimal weighted
one-dimensional \(k\)-means partition and its centres can be found in
\(O(n\log(h+2))\) time and \(O(n)\) space on sorted input, and in
\(O(n\log n)\) time on arbitrarily ordered input. The latter bound is
tight at \(k=n/2\) in the algebraic computation-tree model, even for
unit weights when observations may coincide. The same upper bounds
hold for unweighted one-dimensional \(k\)-medians.
\end{corollary}
\begin{proof}
Sort the observations as \(x_1\le\cdots\le x_n\). For either
objective there is an optimal partition into consecutive groups.
Indeed, take the centres of an optimal solution in sorted order and
assign each observation to a nearest centre, breaking ties toward
the first such centre. The nonempty groups are consecutive. If
fewer than \(k\) remain, split groups until there are exactly \(k\)
and refit their centres. Neither step increases the cost.

Let the edge \(i\to j\) represent the group
\(x_{i+1},\ldots,x_j\). A \(k\)-edge path from \(0\) to \(n\)
partitions the observations into \(k\) such groups. For squared
distances, write \(P_j,Q_j,R_j\) for the prefix sums of \(p_i\),
\(p_ix_i\), and \(p_ix_i^2\) through index \(j\), with zero initial
sums. The best centre of the group is its weighted mean
\((Q_j-Q_i)/(P_j-P_i)\). Substituting it gives the edge weight
\[
 w(i,j)=R_j-R_i-\frac{(Q_j-Q_i)^2}{P_j-P_i}.
\]
The prefix sums take \(O(n)\) time to build and give each weight in
constant time.

To verify~\eqref{eq:monge}, let \(E(X)\), \(p_X\), and \(\mu_X\)
be the optimal cost, total weight, and weighted mean of a nonempty
group \(X\). For consecutive groups \(X,Y\), the cost of joining
them satisfies
\[
 E(X\cup Y)-E(X)-E(Y)
     =\frac{p_Xp_Y}{p_X+p_Y}(\mu_Y-\mu_X)^2 .
\]
If \(Z\) is the next group, replacing \(Y\) by \(Y\cup Z\) increases
its weight and moves its mean weakly to the right of \(\mu_Y\).
Since \(\mu_X\le\mu_Y\), both factors on the right are
nondecreasing under this change. It follows that
\[
 E(X\cup Y)+E(Y\cup Z)\le E(X\cup Y\cup Z)+E(Y),
\]
which is~\eqref{eq:monge} at the four group boundaries. The path
in Theorem~\ref{thm:main} gives the cluster boundaries, and the
prefix sums give the centres.

For absolute distances, let \(C(i,j)\) be the cost of the same
interval at its lower median \(x_t\), where
\(t=\lfloor(i+j+1)/2\rfloor\). Prefix sums of the coordinates
evaluate the distances on either side of \(x_t\) in constant time.
For \(j\ge i+2\), \(x_t\) is a median both before and after appending
\(x_j\), so
\[
 C(i,j)-C(i,j-1)=x_j-x_t.
\]
For fixed \(j\) this increment is nonincreasing in \(i\), since
the median index moves to the right. For \(a<b<c<d\), sum the
increments for \(j=c+1,\ldots,d\) at the two starting indices
\(a,b\). This gives \(C(b,d)-C(b,c)\le C(a,d)-C(a,c)\), as
required. Thus Theorem~\ref{thm:main} applies to this objective as
well. For either objective the initial sort, if needed, costs
\(O(n\log n)\) time.

Finally, consider \(m+1\) arbitrary real numbers. Append \(m-1\)
copies of the first, give all \(2m\) observations unit weight, and
ask for \(m\) clusters. The optimum squared error is zero exactly
when two of the original numbers coincide. Indeed, zero cost requires
at most \(m\) distinct values, and any grouping by distinct value
can be split into exactly \(m\) nonempty constant groups. The
construction and the evaluation of a returned partition take
\(O(m)\) time. Ben-Or's \(\Omega(m\log m)\) bound for element
distinctness in the algebraic computation-tree model~\cite{benor}
therefore gives the claimed lower bound.
\end{proof}

\subsection{Maximum-perimeter inscribed polygons}

Given the \(n\) distinct vertices of a convex polygon \(P\) of positive
area in cyclic order, choose \(k\) of them to maximize the perimeter
of the polygon they form. The chosen vertices must also enclose positive
area, and straight angles are allowed. A \emph{rooted} solution must include a specified
input vertex. We use a real-RAM on which arithmetic, Euclidean
distances, and exact comparisons take constant time.

Boyce et al.~\cite{boyce} first compute a maximum-perimeter \(k\)-gon
rooted at one input vertex and then use it to find a global optimum.
Their algorithm takes \(O(nk\log n+n\log^2 n)\) time. Aggarwal et
al.~\cite[Section~6.1]{smawk} reduced the rooted phase to \(O(nk)\)
and the second phase to \(O(n\log n)\). The rooted phase is a fixed-link
shortest-path problem~\cite[Section~3.2]{blp}. Thus the parametric
search of Aggarwal, Schieber, and Tokuyama~\cite{ast} gives
\(O(n\sqrt{k\log n}+n\log n)\) time overall. For \(k=\Omega(\log n)\),
Schieber's algorithm~\cite{schieber} gives
\(n\,2^{O(\sqrt{\log k\,\log\log n})}\) time for the rooted phase.
For integer coordinates in \([1,U]\) with no three vertices collinear,
Bein, Larmore, and Park~\cite[Section~3.2]{blp} obtained
\(O(n(1+\log(nU)))\) time for the global problem.

Our algorithm computes a rooted optimum for the prescribed \(k\) in
\(O(n\log n)\) time or less. Together with the classical second phase,
this gives the same \(O(n\log n)\) global bound for arbitrary real
coordinates that was known for polynomially bounded integer coordinates
in general position.

\begin{corollary}\label{cor:polygon}
For \(3\le k\le n\), a maximum-perimeter \(k\)-gon inscribed in \(P\)
can be found in deterministic \(O(n\log n)\) time. Its vertices can
be chosen from the input. Among \(k\)-gons on the listed vertices
that contain a specified listed root, an optimum can be found in
\(O(n\log(\min\{k-1,n-k\}+2))\) time.
\end{corollary}
\begin{proof}
\emph{A prescribed root.}
Number the listed vertices \(v_0,\ldots,v_{n-1}\) cyclically from the
root \(v_0\), and set \(v_n=v_0\). A \(k\)-edge path from \(0\) to \(n\)
selects \(k\) distinct listed vertices, including the root.
Give \(i\to j\) weight \(-\lVert v_i-v_j\rVert_2\). Its path weight
is the negative of the selected perimeter. For \(a<b<c<d\) with four
distinct points, the segments \(v_av_c\) and \(v_bv_d\) intersect,
including when some vertices are collinear. The triangle inequality
at a point of intersection gives
\[
 \lVert v_a-v_c\rVert_2+\lVert v_b-v_d\rVert_2
 \ \ge\ \lVert v_a-v_d\rVert_2+\lVert v_b-v_c\rVert_2.
\]
If \(a=0\) and \(d=n\), the same inequality is the triangle inequality
through \(v_0\). Thus the edge weights satisfy~\eqref{eq:monge}.

When input vertices lie on a common side, a path could select \(k\)
collinear points. To enforce positive area, consider each maximal
straight side \(S\) containing \(v_0\). There are at most two. The
vertices outside \(S\) form a consecutive interval of indices \([p,q]\)
between the two copies of \(v_0\). Write \([A]\) for \(1\) if
condition \(A\) holds and \(0\) otherwise. For this side, add
\[
 M\,[i<p]\,[j>q]
 \qquad\text{to the weight of }i\to j,\qquad
 M=2\,\operatorname{perimeter}(P).
\]
A path pays this penalty exactly once if it skips the entire interval
and pays zero otherwise. The indicator satisfies~\eqref{eq:monge}
because its first factor is nonincreasing in \(i\) and its second
is nondecreasing in \(j\). The modified weights are therefore Monge.
Three distinct collinear boundary points must lie on a common side,
so a selected polygon incurs a penalty exactly when all its vertices
are collinear. Every selected perimeter is at most
\(\operatorname{perimeter}(P)\) by the triangle inequality. A penalized
path hence has positive weight, whereas every positive-area choice
has negative weight. Such a choice exists for every listed root and
\(3\le k\le n\). One scan computes \(M\) and the at most two intervals,
after which each edge weight takes constant time to evaluate.
Theorem~\ref{thm:main} now gives the rooted bound.

\emph{A global optimum.}
Call a listed vertex a \emph{corner} if its interior angle is strictly
less than \(\pi\). If there are at most \(k\) corners, take them all
and fill the remaining positions with unused listed vertices in cyclic
order. The resulting polygon has the full perimeter of \(P\).

Otherwise an optimum can be chosen entirely from the corners, even
if arbitrary boundary points are allowed. To see this, take any
inscribed \(k\)-gon and slide a vertex in the relative interior of a
side along that side, keeping its neighbours fixed and preserving
cyclic order. The sum of its two incident lengths is convex in its
position, so some endpoint of the permitted segment gives at least
the current perimeter. There the vertex either reaches a corner or
coincides with a neighbour. In the latter case delete the repeated
copy, retaining the cyclic sum of edge lengths even if only two
vertices remain. Each move reduces the number of vertices away from
corners. Insert unused corners in cyclic order until there are \(k\) distinct
vertices. By the triangle inequality, this can only increase the
perimeter or leave it unchanged. The final \(k\) corners enclose
positive area.

Discard all noncorners in one scan, leaving a polygon of \(m>k\)
corners. Choose any corner as the root and compute a rooted optimum
\(Q=(q_1,\ldots,q_k)\) by the first part. The second phase of Boyce
et al.~\cite{boyce}, accelerated by Aggarwal et
al.~\cite[Section~6.1]{smawk}, takes this one rooted polygon and the
cyclic list of corners as input. A global optimum can be numbered
so that its \(i\)th vertex lies on the closed arc between \(q_i\) and
\(q_{i+1}\), with \(q_{k+1}=q_1\). The second phase bisects the arc
of possible roots and uses matrix search on the corresponding subarcs,
with \(O(m)\) work per level and \(O(\log m)\) levels. It returns a
global optimum in a further \(O(m\log m)\) time. Both phases and the
initial scan fit in \(O(n\log n)\) time.
\end{proof}

\subsection{Length-limited binary Huffman coding}

Given \(n\ge2\) symbols of positive weights \(p_1,\ldots,p_n\) and an
integer \(D\), we seek a code in which each codeword has at most
\(D\) bits and no codeword is a prefix of another. Such a code is
represented by a binary tree with one leaf per
symbol. If the leaf for symbol \(i\) is at depth \(\ell_i\), its codeword
is the sequence of edge labels on its root-to-leaf path. The objective is
to minimize \(\sum_i p_i\ell_i\) subject to \(\ell_i\le D\).
A feasible tree exists exactly when \(D\ge\lceil\log_2 n\rceil\).

The package-merge algorithm of Larmore and Hirschberg~\cite{larmoreh}
and the dynamic program of Golin and Zhang~\cite{golinzhang} both use
\(O(nD)\) time and \(O(n)\) space. On sorted input, reverse package-merge
takes \(O(n(D-\log_2 n+1))\) time~\cite{turpinm}, which is linear when
\(D=\lceil\log_2 n\rceil+O(1)\). The reduction to fixed-link shortest paths
also gives \(O(n\sqrt{D\log n}+n\log n)\) time by parametric
search~\cite{ast}, and \(n\,2^{O(\sqrt{\log D\,\log\log n})}\) time
with recursive parametric search~\cite{schieber}. For integer
frequencies, Bein, Larmore and Park gave a bound that depends on their
magnitudes~\cite[Section~3.1]{blp}. An unrestricted Huffman tree can
be constructed in linear time on sorted input and is optimal here
whenever it satisfies the depth bound~\cite{larmoreh}.

Our bound for sorted real weights depends logarithmically on \(D\).
For instance, at \(D=2\lceil\log_2 n\rceil\) it gives \(O(n\log\log n)\)
time, while the package-merge and reverse package-merge bounds are both
\(O(n\log n)\).

\begin{corollary}\label{cor:huffman}
For any feasible \(D\), an optimal depth-limited binary prefix tree
and its codeword lengths can be found in \(O(n\log(D+1))\) time and
\(O(n)\) space when the weights are sorted. Arbitrarily ordered weights
take \(O(n\log n)\) time, including sorting.
\end{corollary}

\begin{proof}
Assume \(p_1\le\cdots\le p_n\) and put \(S_j=\sum_{i=1}^j p_i\), with
\(S_0=0\). An optimal tree is full, since contracting an internal node
with only one child lowers the cost. A full binary tree on \(n\) leaves
has \(n-1\) internal nodes, so we may cap \(D\) at \(n-1\).

\emph{One edge per tree level.}
Place the root at level \(D\) and each child one level below its parent.
Let \(s_j\) count all internal nodes at levels \(0\) through \(j\).
Then \(s_0=0\) and \(s_D=n-1\). If \(s_{j-1}=s\) and \(s_j=t\), the
\(t\) internal nodes have \(2t\) children between them. Of these
children, \(s\) are internal, leaving \(2t-s\) leaves below level \(j\).
For a given shape the smallest weights can be assigned to the deepest
leaves, so these leaves have total weight \(S_{2t-s}\). Summing this
quantity over all levels counts each weight once per edge on its way to
the root.

Accordingly, the dynamic program of Golin and
Zhang~\cite[Eq.~(3) and Appendix~A]{golinzhang} minimizes the cost of
\(D\) transitions from state \(0\) to \(n-1\). A transition
\(s\to t\) with \(s<t\) and \(2t-s\le n\) costs \(S_{2t-s}\).
The transition \(0\to0\) costs zero and allows empty bottom levels
when the tree has height less than \(D\). These are all the allowed
transitions. Every feasible full tree gives such a state sequence when its
leaves are assigned by depth. Conversely, every minimum-cost sequence
encodes a tree~\cite[Appendix~A, Lemma~7]{golinzhang}. This last
property lets us reconstruct a tree from any optimum returned by the
path algorithm.

\emph{Finite Monge weights.}
Use \(D\) consecutive vertices labeled \(0\), followed by one vertex
of each label \(1,\ldots,n-1\). We seek a \(D\)-edge path from the
first zero to \(n-1\). The zero copies allow every possible amount of
initial padding, since a complete sequence has at most \(D\) zero
states. To give every forward edge a finite weight, extend \(S\) by
\[
 F(j)=
 \begin{cases}
    S_j, & 0\le j\le n,\\
    S_n+(j-n)M, & n<j\le 2n-2,
 \end{cases}
 \qquad M=(D+1)S_n .
\]
An edge from label \(s\) to label \(t\) receives weight \(F(2t-s)\).
A feasible tree gives a path of weight at most \(D S_n\). An edge
with \(2t-s>n\) alone costs at least \(S_n+M>D S_n\), and all weights
are nonnegative. Hence an optimal path uses only allowed transitions.

The successive differences of \(F\) are nondecreasing. For any four
vertices in order, write their labels as \(a\le b\le c\le d\).
The difference \(F(2c-a)-F(2c-b)\) sums \(b-a\) successive
differences of \(F\). Shifting this interval right by \(2(d-c)\)
can only increase the sum. Thus
\[
 F(2c-a)-F(2c-b)\le F(2d-a)-F(2d-b),
\]
which is precisely~\eqref{eq:monge} for the four vertices, including
the case of repeated zero labels.

\emph{Recovering the tree.}
Read the state labels \(s_0,\ldots,s_D\) from an optimal path.
Put \(a_0=0\) and \(a_j=s_j-s_{j-1}\) for \(1\le j\le D\), so
\(a_j\) is the number of internal nodes on level \(j\). Proceed from
the bottom up. On level \(j\), create \(a_j\) parents. Their \(2a_j\)
child positions take the \(a_{j-1}\) existing internal nodes on
level \(j-1\) and \(2a_j-a_{j-1}\) new leaves. Assign the next
smallest weights to these leaves, in any order among ties. This
constructs all \(2n-1\) nodes in \(O(n)\) time. A traversal gives
the depth, and hence the codeword length, of each leaf in the same
bound.

There are \(n+D-1=O(n)\) graph vertices. The prefix sums take \(O(n)\)
time to compute and give constant-time edge queries. With \(D\) links,
the parameter \(h\) in Theorem~\ref{thm:main} is
\(\min\{D-1,n-2\}=D-1\). The theorem and the reconstruction therefore
give the stated time and space bounds.
\end{proof}

\paragraph{AI disclosure.}
We used GPT Astra 6.2 interactively to discover the algorithm, develop
its proofs, and improve the exposition. The authors take full
responsibility for the content of this paper.

\bibliographystyle{abbrvnat}
\bibliography{references}

\end{document}